\documentclass[aps,prx,twocolumn,showpacs,groupedaddress,amsmath,amssymb]{revtex4-1}
\usepackage{graphicx}
\usepackage{dcolumn}
\usepackage{bm}
\usepackage{subfigure}
\usepackage[rgb]{xcolor}
\newcommand{\beginsupplement}{%
        \setcounter{table}{0}
        \renewcommand{\thetable}{S\arabic{table}}%
        \setcounter{figure}{0}
        \renewcommand{\thefigure}{S\arabic{figure}}%
     }

\begin{document}

\title{Using RIXS to uncover elementary charge and spin excitations in correlated materials}

\author{Chunjing Jia$^{*1}$,  Krzysztof Wohlfeld$^{*1, 2}$, Yao Wang$^{1, 3}$, Brian Moritz$^{1}$, Thomas P. Devereaux$^{1, 4}$
}
\affiliation{$^{1}$ Stanford Institute for Materials and Energy Sciences, SLAC National Laboratory and Stanford University, Menlo Park, CA 94025, USA}
\affiliation{$^{2}$ Institute of Theoretical Physics, Faculty of Physics, University of Warsaw, Pasteura 5, PL-02093 Warsaw, Poland}
\thanks{C. J. J. and K. W. contributed equally to this work. \\ Corresponding authors: krzysztof.wohlfeld@fuw.edu.pl and chunjing@stanford.edu}
\affiliation{$^{3}$ Department of Applied Physics, Stanford University, Stanford, CA 94305, USA}
\affiliation{$^{4}$ Geballe Laboratory for Advanced Materials, Stanford University, Stanford, CA 94305, USA}
\date{\today}
\date{\today}

\begin{abstract}
Despite significant progress in 
resonant inelastic x-ray scattering (RIXS) experiments on cuprates at the Cu $L$-edge, 
a theoretical understanding of 
the cross-section remains incomplete 
in terms of elementary excitations 
and the connection to both charge and spin structure factors.  Here we use state-of-the-art, 
unbiased numerical calculations 
to study 
the low energy excitations 
probed by RIXS in undoped and doped Hubbard model relevant to the cuprates.  The results 
highlight the importance of scattering geometry, in particular both the incident and scattered x-ray photon polarization, and demonstrate 
that on a qualitative level 
the RIXS spectral shape in the cross-polarized channel 
approximates that of the spin dynamical structure factor. 
However, in the parallel-polarized channel the complexity of the RIXS process beyond a simple two-particle response complicates the analysis, and demonstrates 
that 
approximations and expansions which attempt to 
relate 
RIXS to 
less complex 
correlation functions can not reproduce the full diversity of RIXS spectral features.  
\end{abstract}


\maketitle

\section{Introduction}
Resonant inelastic x-ray scattering (RIXS) is an experimental technique in which the transferred energy, momentum and polarization associated with incident and scattered x-ray photons 
can be measured and analyzed to reveal information about 
the elementary excitations of a system~\cite{Ament2011}.  
In recent years RIXS has attracted 
considerable attention 
and 
positioned itself as 
a primary experimental technique 
to probe 
the excitations in correlated 
materials, 
especially transition-metal oxides~\cite{Schlappa2009, Braicovich2009, Guarise2010, Braicovich2010, Schlappa2012, Tacon2011, Dean2013a, Dean2013b, Dean2013c, Tacon2013, Ishii2014, Lee2014, Dean2014, Guarise2014, Minola2015, Wakimoto2015}.  RIXS possesses atomic sensitivity with incoming photons resonantly tuned to a specific atomic absorption edge, making it a particularly unique and powerful tool for characterizing excitations across the Brillouin zone in these materials. 

The direct RIXS process consists of two 
dipole transitions, as shown schematically in Fig.~1 for the Cu {\it L}-edge. 
In the first step an incoming photon excites the ground state by promoting an electron from a filled core shell (Cu $2p$) into the valence shell (Cu $3d$). An intermediate state manifold forms following the charge and spin shake-up which accompany the introduction of a local core-hole potential and this new carrier in the valence shell. In the second step an electron 
from the valence shell, possessing appropriate atomic character, fills the core-hole, 
accompanied by an outgoing photon, which leaves the system in an excited final state.  

Due to the complexity of this process and the presence of the intermediate state, an interpretation of 
RIXS spectra has been 
hindered by an incomplete understanding of how it may be 
related to other, more fundamental, response functions governing spin, charge, lattice and orbital excitations. 
In addition, 
it is well known that polarization plays a key role in understanding selective excitations in Raman spectroscopy~\cite{Devereaux2007}, but a full polarization analysis involving both incident and scattered x-rays has become possible only now in experiment and only shown theoretically to be important for spin-flip excitations~\cite{Jia2014}. One would expect that a more 
complete understanding, at a fundamental level, could be obtained by 
analyzing the full theoretical RIXS cross-section, accounting for the influence of both the incident and
scattered light polarization for a given experimental scattering geometry.

\begin{figure*}[t!]
\includegraphics[width=1.8\columnwidth]{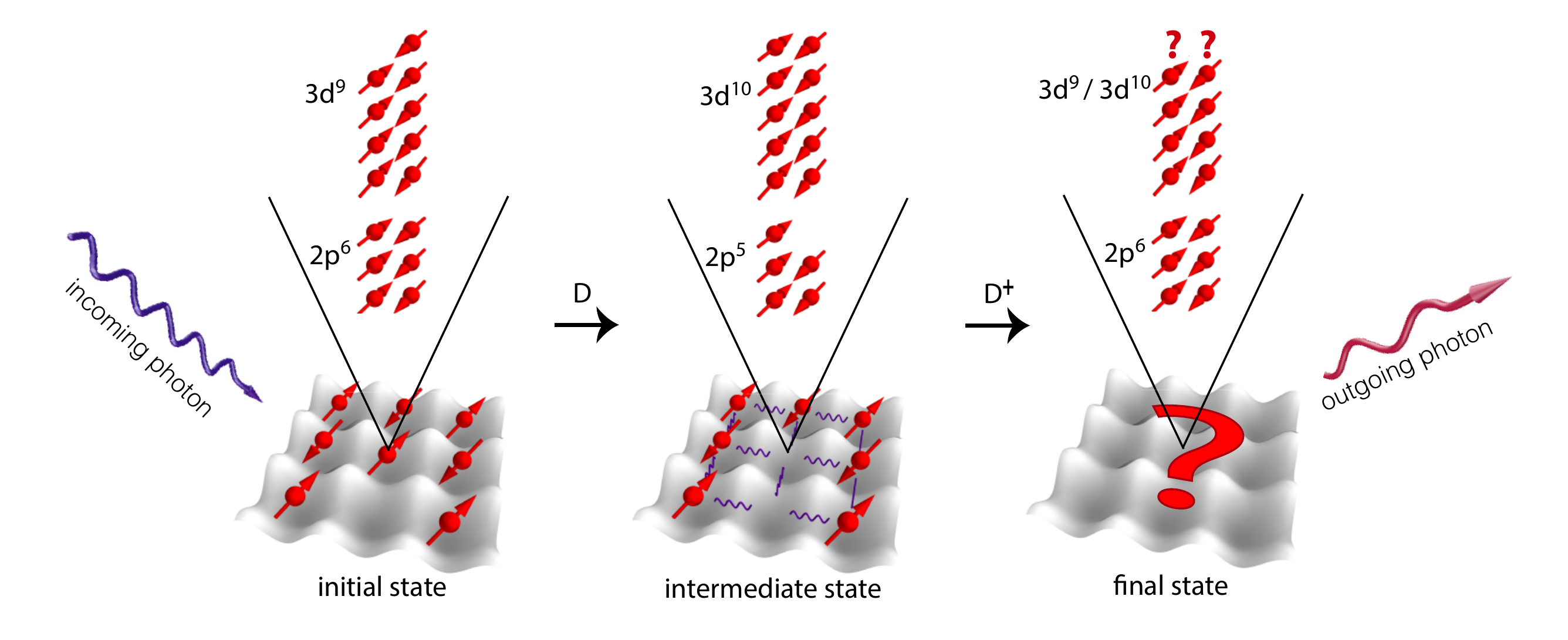}
 \caption{A 
schematic which illustrates the RIXS process at the Cu $L$-edge appropriate for 
cuprates. 
}
\label{fig:cartoon}
\end{figure*}
%


To better understand the RIXS cross-section from a theoretical 
perspective, here we will explore a rather 
general question: 
{\it what 
low energy excitations are measured by RIXS, particularly in cuprates at the Cu $L$-edge?}  There certainly are a number of ways to address this question analytically and numerically. 
Here, our interest is confined to the low energy excitations which involve spin and/or charge degrees of freedom, neglecting both lattice and higher energy orbital or charge transfer excitations in this work.  Our aim will be to qualitatively understand which excitations are encoded in the RIXS cross-section, and any connections to both spin and charge dynamical structure factors, under different scattering conditions; and we will  quantitatively compare numerical estimates of the full RIXS cross-section with various approximations to tease out this information. 

Naively, one may expect the Cu $L$-edge RIXS cross-section to be the resonant analog of the hard x-ray non-resonant IXS,~\cite{Jia2012, Wang2014} whose response is the charge dynamical structure factor.  It then may seem counterintuitive that   
RIXS at the Cu $L$-edge rose to prominence for the successful 
empirical measurement of the spin (magnon) dispersion 
in undoped cuprates, 
making it a complementary experimental probe to the well-established inelastic neutron scattering~\cite{Braicovich2010, Guarise2010}. 
With theoretical support~\cite{Ament2009, Jia2014}, experiments on doped cuprates proved 
sensitive to magnetic excitations, with a similar cross-section to the spin dynamical structure factor, 
regardless of doping level~\cite{Tacon2011, Dean2013a, Dean2013b, Dean2013c, Tacon2013, Ishii2014, Lee2014, Dean2014, Guarise2014, Minola2015, Wakimoto2015}. These observations were interpreted in terms of persistent magnetic excitations up to an unexpectedly high doping level.  However, this semi-empirical connection to the spin dynamical structure factor has been based primarily on approximate theoretical and numerical treatments for the full RIXS cross-section. 
The fast collision approximation~\cite{Luo1993, Groot1998, Veenendaal2006} and the effective operator approach~\cite{Haverkort2010} 
suggested that 
only single magnon excitations or $S(\mathbf{q}, \omega)$ should be measured by RIXS at the Cu $L$-edge~\cite{Ament2009, Haverkort2010, Marra2012}. More sophisticated treatments -- the ultrashort core-hole lifetime (UCL) expansion~\cite{Ament2007, Bisogni2012a, Bisogni2012b} and 
the UCL-inspired ansatz~\cite{Igarashi2012a} -- 
highlighted that RIXS should be sensitive also to bimagnon excitations. 
A further extension of UCL also 
pointed out the importance of 
three-magnon excitations~\cite{Ament2010}. 
However, 
these studies made no explicit comparison between the approximations 
and the full RIXS cross-section, nor have they concentrated on the sensitivity of RIXS to the charge dynamical structure factor. 
  
Perhaps more importantly, 
these approximations 
suffer from several severe limitations. First, the fast collision approximation (or the effective operator approach)~\cite{Ament2009, Haverkort2010, Marra2012} relies on 
an estimation of the dynamics only at the site where the core-hole has been created in the intermediate state.   However, 
the intermediate state 
is drawn from a manifold of states, which differ 
from the ground state not only locally, at the site where a core-hole is created, but also on 
neighboring sites due to 
hopping or spin exchange associated with the dynamical screening process. As a result, and especially upon doping, this approximation 
fails to capture key elements of the full RIXS process. 
Second, the UCL approximation relies on the assumption 
that the energies of the intermediate state manifold 
are 
much 
smaller than the inverse core-hole lifetime $\Gamma$, which 
allows an approximation based on only the first few (two) terms in the UCL Taylor-series-like expansion. However, this assumption 
need 
not hold, 
especially in 
``itinerant'', doped systems 
where a number of intermediate states 
may have energies $\propto t \sim \Gamma $, making 
the UCL 
a non-convergent approximation. 


In this paper we 
study the low energy excitations 
of RIXS at the Cu $L$-edge 
in an unambiguous way by numerically evaluating the cross-section 
for a two-dimensional Hubbard model comprising ``effective'' Cu $3d$ orbitals supplemented by 
local Cu $2p$ core levels. 
Using 
exact diagonalization, 
which previously was 
applied to the study of paramagnons in cuprates~\cite{Jia2014}, 
we 
compare the 
exact RIXS cross-sections to 
approximate 
cross-sections 
for the same model 
using the same method. 
The intrinsically correlated nature of our Hamiltonian distinguishes 
this study from the analytically exact RIXS calculations for the case of completely uncorrelated electrons~\cite{Benjamin2014}.
In the next section, we present and compare numerical results for the 
exact and approximate RIXS cross-sections, 
and discuss 
content of the excitations 
and 
consequences 
for 
experiments. The paper ends with 
conclusions, and appendices which contain details and longer derivations. 

\section{Numerical results}
\label{sec:numerics}

\subsection{Exact RIXS cross-section}
\label{sec:exact}
 
The RIXS cross-section at the Cu $L$-edge is~\cite{Ament2011, Marra2012}
%
\begin{equation}
\label{eq:crosssection}
I_{\bf e}({\bf q}, \omega)= \sum_f \Big| \Big\langle f \Big\vert O_{{\bf q},{\bf e}} \Big\vert i \Big\rangle \Big|^2 \delta (\omega + E_i - E_f),
\end{equation}
where $\vert i \rangle $ ($\vert f \rangle $) is the initial (final) state of the system in the RIXS process with energy $E_i$ ($E_f$), 
transfered 
momentum (energy loss) is ${\bf q}\equiv{\bf k}_{ i} - {\bf k}_{f} $ ($\omega \equiv {\omega}_{ i} - {\omega}_{ f} $) where ${\bf q}_{ i}$ and ${\bf q}_{f}$ (${\omega}_{ i}$ and ${\omega}_{ f}$) are the incoming and outgoing photon momentum (energy), and $\mathbf{e}=\mathbf{e}^{i}\cdot(\mathbf{e}^{f})^{\dagger}$ is the tensor that describes the incoming ($i$) and outgoing ($f$) photon polarizations. Here the operator $O_{{\bf q},{\bf e}} =  1/\sqrt{N} \sum_{\bf j}e^{i{\bf q\cdot j}}  O_{{\bf j},{\bf e}}$ and
\begin{equation}
\label{eq:operator}
{O}_{{\bf j}, {\bf e}} = D_{{\bf j}, {\bf e}^{f}}^\dag \frac{1}{\omega_{i}-\mathcal{H}+\imath\Gamma} D_{{\bf j}, {\bf e}^{i}},
\end{equation}
which describes the evolution of the system in the RIXS experiment from the initial state to the final state via the intermediate states accessible 
via the core-hole to valence band dipole transitions. $N$ is the number of lattice sites in the system. The dipole transition operator $D_{{\bf j}, {\bf e}} = \sum_{\sigma, \alpha, \beta} (A^{\bf e}_{\alpha} p^\dag_{{\bf j} \alpha \sigma} d_{{\bf j} \sigma} + h.c.)$ 
with $p_{{\bf j} \alpha \sigma}$ ($d_{{\bf j} \sigma}$) annihilates a hole in the $2p$ ($3d$) shell with spin $\sigma$.
$A^{\bf e}_{\alpha}$ is the matrix element of the dipole transition between the 2$p_{\alpha}$ orbitals and 3$d_{x^2-y^2}$ orbital written as $A^{\bf e}_{\alpha} = \langle d_{x^2-y^2, \sigma} | \hat{\epsilon} \cdot \hat{r} | p_{\alpha\sigma} \rangle$ for polarization $\hat{\epsilon}$. $\Gamma$ is the inverse core-hole lifetime.
Note that with the exception of the schematic in 
Fig.~\ref{fig:cartoon}, ``hole notation'' has been used throughout the paper, 
such that the dipole transitions at the Cu $L$-edge correspond 
to transitions from the initial $3d^{*1} 2p^0$ state to a $3d^{*0} 2p^1$ intermediate state, and finally from the intermediate state to the final 
 state configuration. Here $3d^{*}$ corresponds to holes nominally (Zhang-Rice singlets~\cite{ZRS}) in the single band notation. 

Here we use the single-band Hubbard model as it carries the key features of correlated materials in the low energy regime, which are the relevant energy regime for the study of charge and spin excitations for cuprates in RIXS measurements. The single-band results can be applied directly to cuprates and can be generalized to other multi-orbital correlated materials. The Hamiltonian $\mathcal{H}$ defined on a 2D square lattice describes the relevant interactions of the ``effective'' $3d$ and $2p$ orbitals, 
consisting 
of two parts:
\begin{align}
\label{eq:hamiltonian} 
\mathcal{H}&= H+H_c, \\
{{H}}&\!=\! -t\!\! \sum_{\langle {\bf  i}, {\bf j} \rangle,\sigma} d_{{\bf i} \sigma}^{\dagger}d_{{\bf j} \sigma}\!-\!t^{\prime}\!\! \sum_{\langle \langle {\bf i}, {\bf  j} \rangle \rangle,\sigma} d_{{\bf i} \sigma}^{\dagger}d_{{\bf j} \sigma}\!+\!U\!\sum_{\bf i} n_{{\bf i} \uparrow}^d n_{{\bf i} \downarrow}^d,  \\
{{H}}_{c}&\!=\! (\epsilon^d\!-\!\epsilon^p) \sum_{{\bf i} \alpha\sigma} n_{{\bf i} \alpha\sigma}^p
+\mathit{U}_c\sum_{{\bf i} \alpha\sigma\sigma^{\prime}} n_{{\bf i} \sigma}^d n_{{\bf i} \alpha\sigma^{\prime}}^p \nonumber \\
&+\lambda\sum_{{\bf i} \alpha\alpha^{\prime} \sigma\sigma^{\prime}} p_{{\bf i} \alpha\sigma} ^{\dagger} \chi_{\alpha\alpha^{\prime}}^{\sigma\sigma^{\prime}} p_{{\bf i} \alpha^{\prime}\sigma^{\prime}}.
\end{align}
The first part, Eq.~(4), is the well-known single-band Hubbard model with the nearest (next-nearest) neighbor hopping $t$ ($t'$) and on-site Hubbard repulsion $U$. Here, the operator $d^{\dagger}$ in the single-band Hubbard model creates a ``$3d^{*}$'' hole, which should be distinguished from the actual Cu $3d_{x^2-y^2}$ hole in the multiband model.  The second part, Eq.~(5), describes (i) the energy splitting between the $2p$ and the $3d$ shells through the difference in site energy $\epsilon^d\!-\!\epsilon^p$, (ii) the repulsion between the $2p$ and the $3d$ holes (the ``core-hole potential'') $U_c$,
and (iii) the spin-orbit coupling in the $2p$ shell $\lambda$ with the matrix elements $\chi_{\alpha\alpha^{\prime}}^{\sigma\sigma^{\prime}}\equiv\langle p_{\alpha\sigma}|\mathbf{l}\cdot \mathbf{s} |p_{\alpha^{\prime}\sigma^{\prime}}\rangle$, where the $\mathbf{l}\cdot \mathbf{s}$ term represents the spin-orbit coupling operator. 

\begin{figure*}[t!]
\includegraphics[width=1.8\columnwidth]{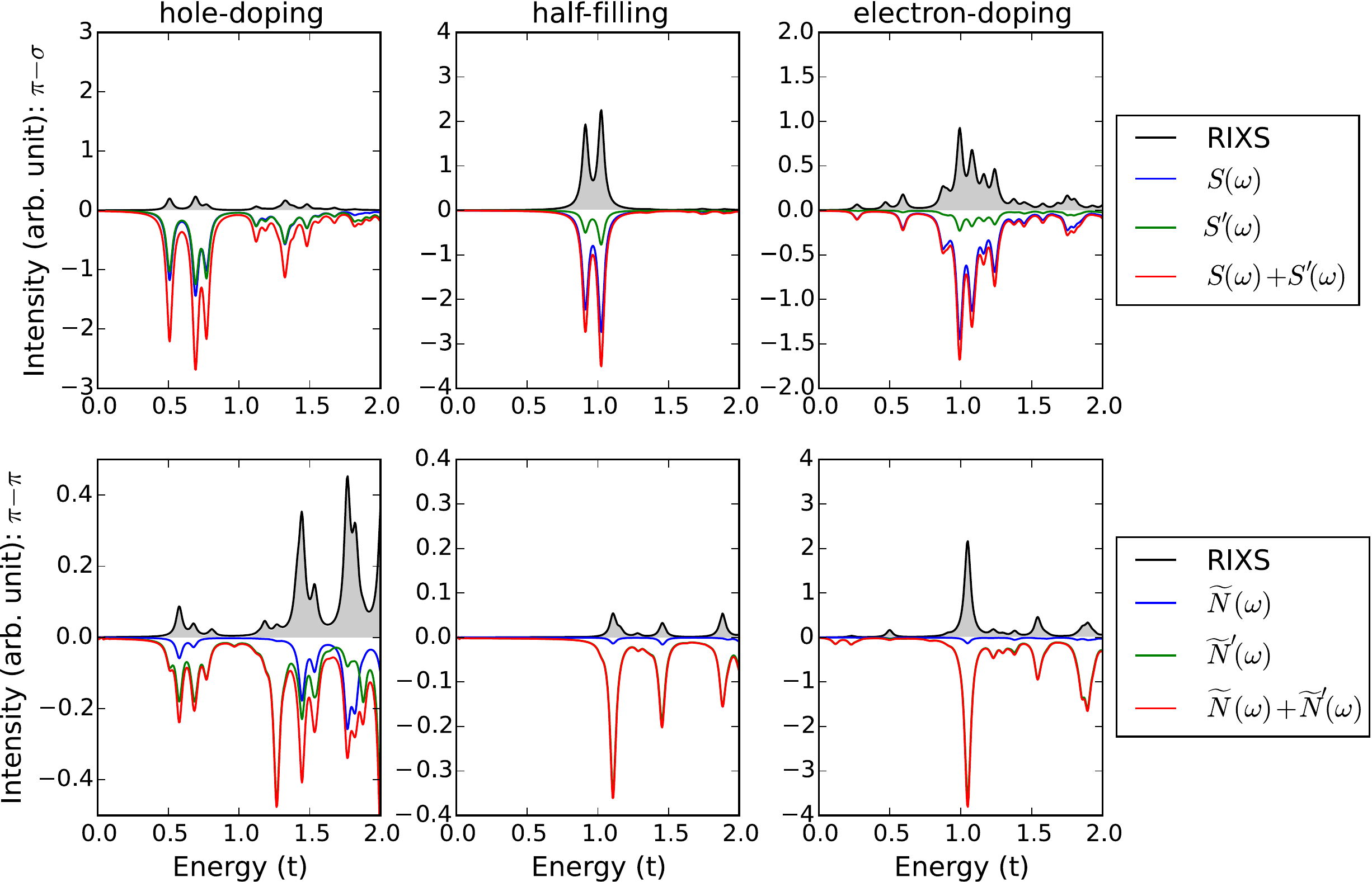}
 \caption{Exact RIXS cross-sections and approximations calculated using exact diagonalization. 
 Top (bottom) panels show spectra for the  $\pi-\sigma$ ($\pi-\pi$) polarization 
 geometries.  Left, middle, and right panels show RIXS spectra calculated for $n=0.83$ (``hole-doping''), $n=1$ (``half-filling''), and $n=1.17$ (``electron doping'') electron fillings, respectively. The spectra are summed over all momenta along the nodal direction and along the $k_y=0$ direction that are
accessible in RIXS 
on a 12-site cluster, 
weighted with the RIXS form factors, i.e.: $S(\omega)= \sum_{\bf q}  |{W}_{\bf \pi\!-\!\sigma} |^2 S({\bf q}, \omega) $, $\tilde{N} (\omega)= \sum_{\bf q}  |{W}_{\bf \pi\!-\!\pi} |^2 \tilde{N} ({\bf q}, \omega)$
and similarly for $S' (\omega)$ and  $\tilde{N'}(\omega)$. The elastic response has been removed in each panel.
}
\label{fig:proxy}
\end{figure*}

The RIXS cross-section is calculated using exact diagonalization on a 12-site cluster. In this cluster, momentum points (2$\pi$/3,0) and ($\pi$/2, $\pi$/2) are accessible, providing information along both the Brillouin zone axis and diagonal, relevant for RIXS experiments. As the Cu 2p core levels and the spin-orbit coupling are included (which also means that the total spin is not a good quantum number in the intermediate state), the Hilbert space size is $\sim 10^{7}$. Ground state eigenvectors and eigenvalues were obtained using the implicitly restarted Arnoldi method 
encoded in the Parallel ARPACK~\cite{Lehoucq1998} libraries. 
The 
cross-section itself was 
obtained using the biconjugate gradient stabilized method~\cite{Vorst1992} and continued fraction expansion~\cite{Dagotto1994}. The numerical technique has been applied previously to 
calculate 
RIXS at the Cu $K$-edge and $L$-edges~\cite{Jia2012, Jia2014}. 
Numerical results were obtained for parameters 
which can relatively well reproduce the low energy physics for cuprates: 
$\epsilon^d-\epsilon^p=2325 t$ (which gives the typical splitting between the Cu $2p$ and the Cu $3d$ shell
of 930 eV if $t=0.4$eV), $U_c = 4t$, $\lambda = 32.5 t$, $\Gamma=t$, $U = 8t$, $t^{\prime}=-0.3t$. RIXS spectra at half-filling are taken at the Cu $L_3$ resonance (i.e. $\omega_i \sim \epsilon^d\!-\!\epsilon^p + E_{L_3}$ with $E_{L_3} = - \lambda / 2  $). 
For doped 
systems, we investigate at the resonance where the character of 
the intermediate state 
is similar to that 
in the undoped case on the core-hole site. The angle between the incident and the scattered photons 
is set to 
$50^\circ$, with the scattering plane parallel to $xz$, i.e. perpendicular to the $xy$ plane on which we define the 2D Hubbard Hamiltonian, and the incoming polarization is chosen to be $\pi$. This scattering geometry 
is consistent with that 
used most commonly 
in RIXS 
measurments 
for cuprates~\cite{Dean2013a, Dean2013b, Dean2013c, Ishii2014, Lee2014, Dean2014, 
Guarise2014}. 
The relation between polarization 
and the transferred momenta~\cite{Veenendaal2006, Ament2009, Haverkort2010, Wohlfeld2013} follows from this scattering geometry: ${\bf e}^{i} = (\sin \theta, 0, \cos \theta)$,  ${\bf e}^{f} = [-\cos (\theta-40^\circ), 0, \sin(\theta-40^\circ)]$ [${\bf e}^{f} = (0, -1, 0)$] for outgoing $\pi$ ($\sigma$) polarization, $\theta \in [0^\circ, 130^\circ]$, and the angle $\theta$ is related to the transferred momentum via $k_x = 1.07 \pi \sin (\theta - 65^\circ)$. The calculated RIXS spectra will be presented at Figs.~\ref{fig:proxy} - \ref{fig:spectrum_pi2}. The comparison with the approximated spectra will be presented in Section II C below.

\subsection{Approximate cross-section}  
\label{sec:approximate1}
The approximations follow from integrating 
out the core hole degrees of freedom using two approaches. We firstly assume that the energy of the incoming photon $\omega_i$ is tuned to the main resonance at the Cu $L_3$ edge, as dictated by $H_c$. \footnote{At this resonance, the intermediate state 
is an element of 
a subspace with no $3d^{*}$ hole on the core-hole site.} This is equivalent to a projection 
into a subspace with only one $3d^{*}$ hole on the core-hole site in either the initial state $|i\rangle$ or the final state $|f\rangle$. We then 
perform the UCL approximation by expanding the RIXS operator, Eq.~(\ref{eq:operator}), in a power series in $\mathcal{H}/\Gamma$ and only keeping the first two terms. See Appendix A for details. 
These approximations can be written separately for the two polarization conditions as follows:

\begin{figure}[t!]
\includegraphics[width=1.0\columnwidth]{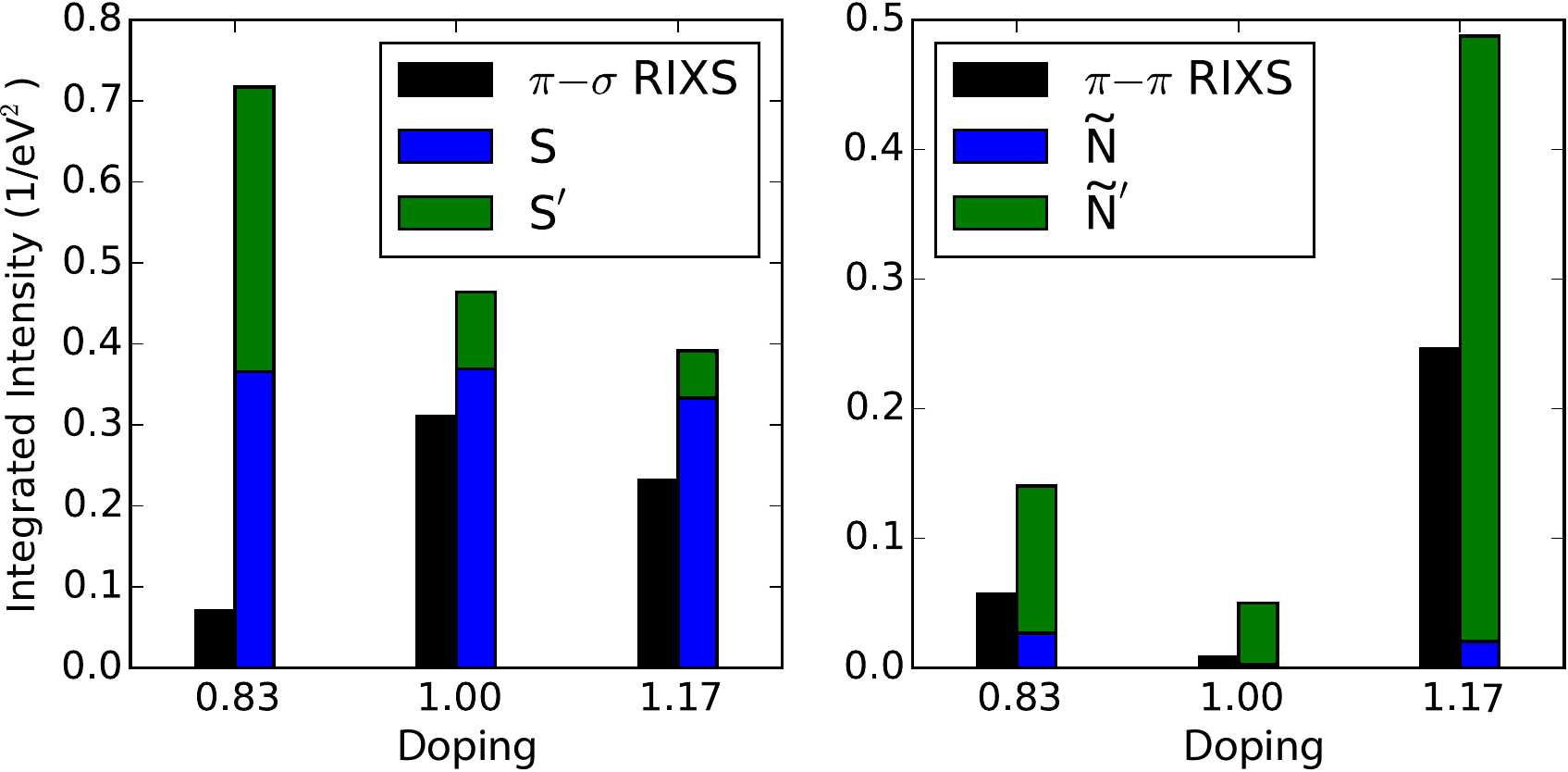}
 \caption{Momentum and energy integrated spectral weight
for full RIXS $\int_{0}^{1.6t} I(\omega) {\rm d}\omega$
and the various approximations $S=\int_{0}^{1.6t} S(\omega){\rm d}\omega$,  $\tilde{N}=\int_{0}^{1.6t} \tilde{N} (\omega){\rm d}\omega$, ${S'}=\int_{0}^{1.6t} S' (\omega){\rm d}\omega$, $\tilde{N'}=\int_{0}^{1.6t} \tilde{N'}(\omega){\rm d}\omega$ with $S(\omega), \tilde{N} (\omega), S' (\omega), \tilde{N'}(\omega)$ defined as in the caption of Fig.~\ref{fig:proxy} RIXS spectra. The $\pi-\sigma$ ($\pi-\pi$) relative polarizations 
are shown on the left (right) panels, respectively.
Note that the spectral weight $\tilde{N}= 5 \times 10^{-5} (eV)^{-2}$ in the half-filled case and does not appear in the figure.
All momenta along the nodal direction and along the $q_y=0$ direction 
accessible on a 12-site cluster 
are taken into account 
(elastic response 
excluded).
}
\label{fig:spectralweight}
\end{figure}

{\it $\pi\!-\!\sigma$ RIXS in the UCL approximation:}
\begin{align}\label{eq:uclpisigma}
 I^{\rm UCL}_{\bf \pi\!-\!\sigma}({\bf q}, \omega)\! =&    |{W}_{\bf \pi\!-\!\sigma} |^2   \Big\{ S({\bf q}, \omega) + S' ({\bf q}, \omega)  \Big\},   \\
 S({\bf q}, \omega)\! =& \sum_f \Big| \Big\langle f \Big\vert  {S}_{\bf q}^z \Big\vert i \Big\rangle \Big|^2 \delta (\omega\! +\! E_i\! -\! E_f),   \\ 
S' ({\bf q}, \omega)\! =& 
\frac{z^2 t^2 }{N^2 \Gamma^2} \sum_f \Big| \Big\langle f \Big\vert   \sum_{{\bf k}, {\bf k'}}  
 \varepsilon_{-{\bf k'}+{\bf k}+{\bf q}}
S^z_{\bf k'} \nonumber \\
& \label{eq:tildeS}
\times {d}_{{\bf k}, \sigma}^\dag 
{d}_{-{\bf k'}+{\bf k}+{\bf q}, \sigma} \Big\vert i \Big\rangle \Big|^2 
\delta (\omega \!+\! E_i\! -\! E_f),
\end{align}
where the local RIXS form factor ${W}_{\bf \pi\!-\!\sigma}  =- \imath 2 \sin \theta / (3 \Gamma)$,
the spin operator ${S}^z_{\bf q} =  1/({2\sqrt{N}}) \sum_{{\bf k}} ( {d}_{{\bf k}, \uparrow}^\dag {d}_{{\bf q}+{\bf k}, \uparrow} - {d}_{{\bf k}, \downarrow}^\dag {d}_{{\bf q}+{\bf k}, \downarrow}) $, $z= 4$ is the 2D
coordination number, and the 2D structure factor is $\varepsilon_{{\bf k}} =  \gamma_{{\bf k}} + t' \eta_{{\bf k}} / t$ with $\gamma_{\bf k} = (\cos k_x + \cos k_y)/2$ and $\eta_{\bf k} = \cos k_x \cos k_y$. The structure factor $\varepsilon_{{\bf k}}$ has $A_{1g}$ symmetry.  
Note that the first term of the expansion $S({\bf q}, \omega)$
has the form of the spin dynamical structure factor, while the second term $S' ({\bf q}, \omega)$ is a rather complicated four particle response which probes 
spin and charge excitations. The latter 
corresponds to the 3-spin Greens function (see Appendix B) {\it and} to 
joint spin and charge excitations. 

{\it $\pi\!-\!\pi$ RIXS in UCL approximation:}
\begin{align}\label{eq:uclpipi}
 I^{\rm UCL}_{\bf \pi\!-\!\pi}({\bf q}, \omega)\! =&    |{W}_{\bf \pi\!-\!\pi} |^2   \Big\{ \tilde{N} ({\bf q}, \omega) + \tilde{N'} ({\bf q}, \omega)  \Big\},   \\
 \tilde{N} ({\bf q}, \omega)\! =& \sum_f \Big| \Big\langle f \Big\vert  \tilde{n}_{\bf q} \Big\vert i \Big\rangle \Big|^2 \delta (\omega \!+\! E_i\! -\! E_f),   \\ 
\tilde{N'} ({\bf q}, \omega)\! =& 
 \label{eq:tildeN}
\frac{z^2 t^2 }{N \Gamma^2} \sum_f \Big| \Big\langle f \Big\vert  \sum_{\bf k} \varepsilon_{{\bf q}+{\bf k}} \tilde{d}_{{\bf k}, \sigma}^\dag 
{d}_{{\bf q}+{\bf k}, \sigma} \Big\vert i \Big\rangle \Big|^2 \nonumber \\
& \times \delta (\omega\! + \!E_i \!-\! E_f),
\end{align}
where the constrained density operator is 
$ \tilde{{n}}_{\bf i} =  \sum_{\sigma} \tilde{d}^{\dag}_{{\bf i} \sigma} \tilde{d}_{{\bf i} \sigma}$ with 
constrained fermions $\tilde{d}^{\dag}_{{\bf i}\sigma} =  {d}_{{\bf i}\sigma}^{\dag}(1-n_{{\bf i} \bar{\sigma}})$
and $\tilde{d}_{{\bf i}\sigma} =  (1-n_{{\bf i} \bar{\sigma}}){d}_{{\bf i}\sigma}$, and the local RIXS form factor ${W}_{\bf \pi\!-\!\pi}  =-  2 \sin \theta \cos (\theta - 40^\circ) / (3 \Gamma)$.

Before evaluating the correlation functions, we want to highlight that {\it none} of these approximations give the standard charge dynamical structure factors for the parallel-polarization channel. The first term of the expansion, $\tilde{N} ({\bf q}, \omega)$,
is {\it not} the standard charge dynamical structure factor. 
It represents a more complicated four-particle response function 
in the original space of unprojected 
fermions. 
In the following, 
it carries information about 
projected charge excitations, which 
should in no way be confused with 
information about the full charge response. 
That RIXS does not probe the 
standard charge response 
stems from the fact that initial states 
with double occupancy on the core-hole site 
can not be excited in the RIXS process 
due to the Pauli principle. The second term $\tilde{N'} ({\bf q}, \omega)$, a complicated four-particle response function as well, 
corresponds to 
2-spin or bimagnon excitations (see Appendix B) 
{\it and} other symmetry projected 
charge excitations, represented again by a correlator beyond the familiar two-particle charge response.

\subsection{Comparing exact and approximate results} 
\label{sec:comparison}
In the following, we will present a systematic comparison of the full RIXS spectra with the approximations using the UCL expansion for the two polarization conditions. As a momentum resolved technique, RIXS has the power to measure the dispersion of elementary excitations, which is one of its main advantages compared to traditional optical or Raman scattering, where the momentum transfer is limited to $q\sim0$.~\cite{Devereaux2007} However, there are limitations in cluster size, as well as a limited number of poles using a finite-size cluster. Thus, summing over all the accessible momentum points (results as shown in Fig.~\ref{fig:proxy}) gives us a complete picture of the energies and the distribution of intensities for the excitations, so that the comparisons to the approximations can be made in a single shot. To better quantify this comparison, we also calculate and compare the total spectral weight carried by the excitations, cf. Fig.~\ref{fig:spectralweight}. Nevertheless, in section II D, we compare exact RIXS cross-sections and approximations at the momentum points (2$\pi$/3,0) and ($\pi$/2, $\pi$/2), and discuss connections to experiments.

The 
full 
cross-sections are shown in Fig.~\ref{fig:proxy} for two polarization channels: (i) the cross-polarized channel [$\pi-\sigma$ with $\pi$ ($\sigma$) incoming (outgoing) polarization] and (ii) the parallel-polarized channel [$\pi-\pi$ with $\pi$ ($\pi$) incoming (outgoing) polarization]. For each 
the spectra are calculated for three different doping levels ($n=0.83$, ``hole-doping'', 
$n=1$, ``half-filling'', and $n=1.17$, ``electron doping''). 
The $\pi-\sigma$ RIXS spectra presented here agree with those presented in Ref.~[\onlinecite{Jia2014}]. The 
approximate 
cross-sections 
$S(\omega)$ and $S^{\prime}(\omega)$ for cross-polarization and $\tilde{N}(\omega)$ and $\tilde{N}^{\prime}(\omega)$ for parallel-polarization 
are also shown in Fig.~\ref{fig:proxy}. The results 
for both full and approximated RIXS are shown for a Lorentzian broadening with half width at half maximum (HWHM) = $0.025t$ for the energy transfer. Note that 
the spectra in Fig.~\ref{fig:proxy} correspond to 
a momentum summation over all points accessible in the 12-site cluster, to provide a holistic picture of the character of excitations probed by RIXS and the utility of various approximations (see Fig.~\ref{fig:proxy} caption for details). 

\begin{figure*}[t!]
\includegraphics[width=1.8\columnwidth]{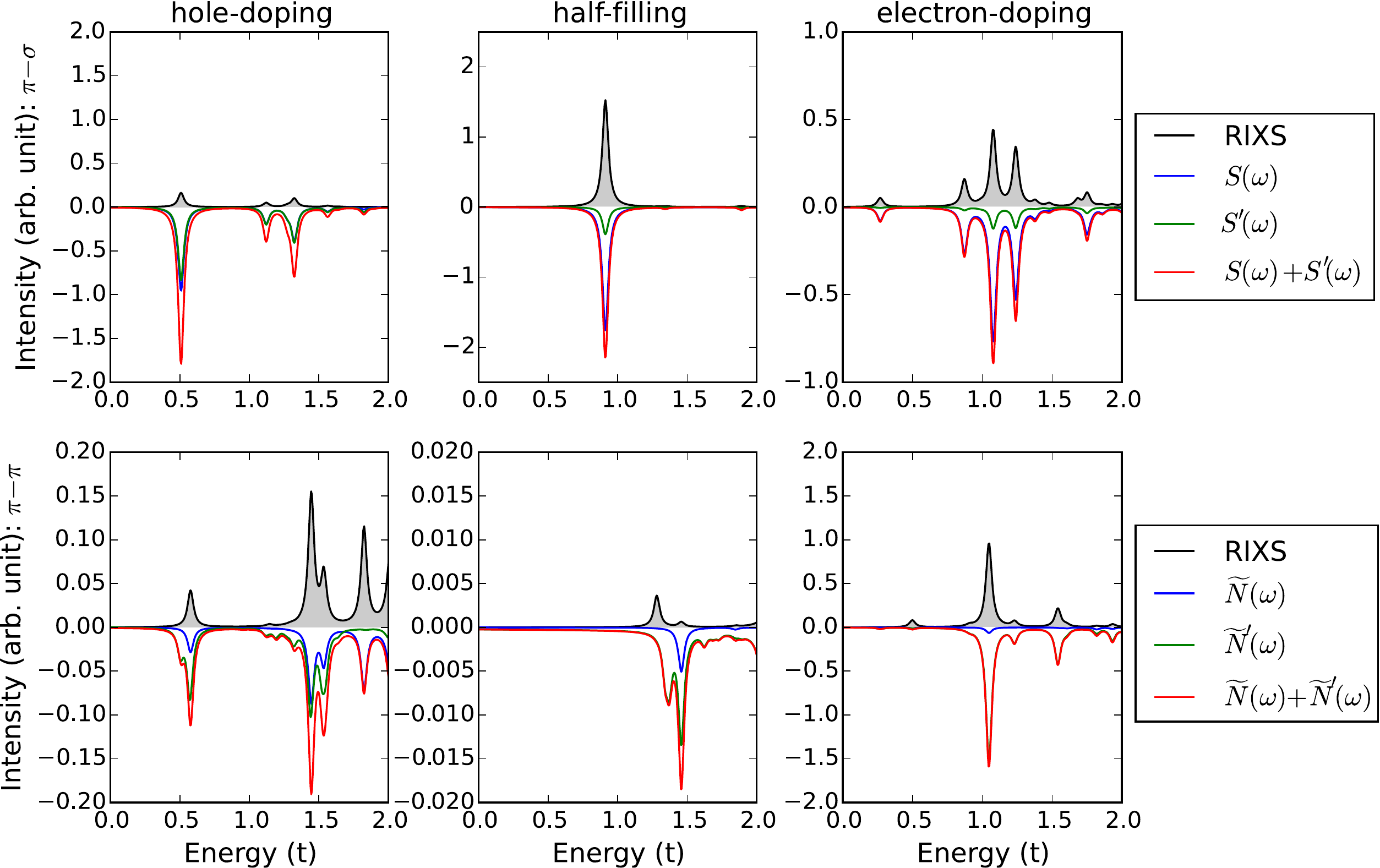}
 \caption{
Cross-sections at {\bf 
${\bf q} = (2\pi/3, 0)$ 
} with the elastic response removed.
Top (bottom) panels show spectra for the  $\pi-\sigma$ ($\pi-\pi$) polarization. 
Left, middle, and right panels show RIXS spectra 
for $n=0.83$, 
$n=1$, 
and $n=1.17$ 
electron fillings, respectively. The approximate spectra are weighted with the RIXS form factors, i.e.
$S({\bf q}, \omega) \rightarrow |{W}_{\bf \pi\!-\!\sigma} |^2 S({\bf q}, \omega) $, $\tilde{N} ({\bf q}, \omega) \rightarrow |{W}_{\bf \pi\!-\!\pi} |^2 \tilde{N} ({\bf q}, \omega)$
and similarly for $S' ({\bf q}, \omega)$ and  $\tilde{N'}({\bf q}, \omega)$.
}
\label{fig:spectrum_2pi3}
\end{figure*}
%

\begin{figure*}[t!]
\includegraphics[width=1.8\columnwidth]{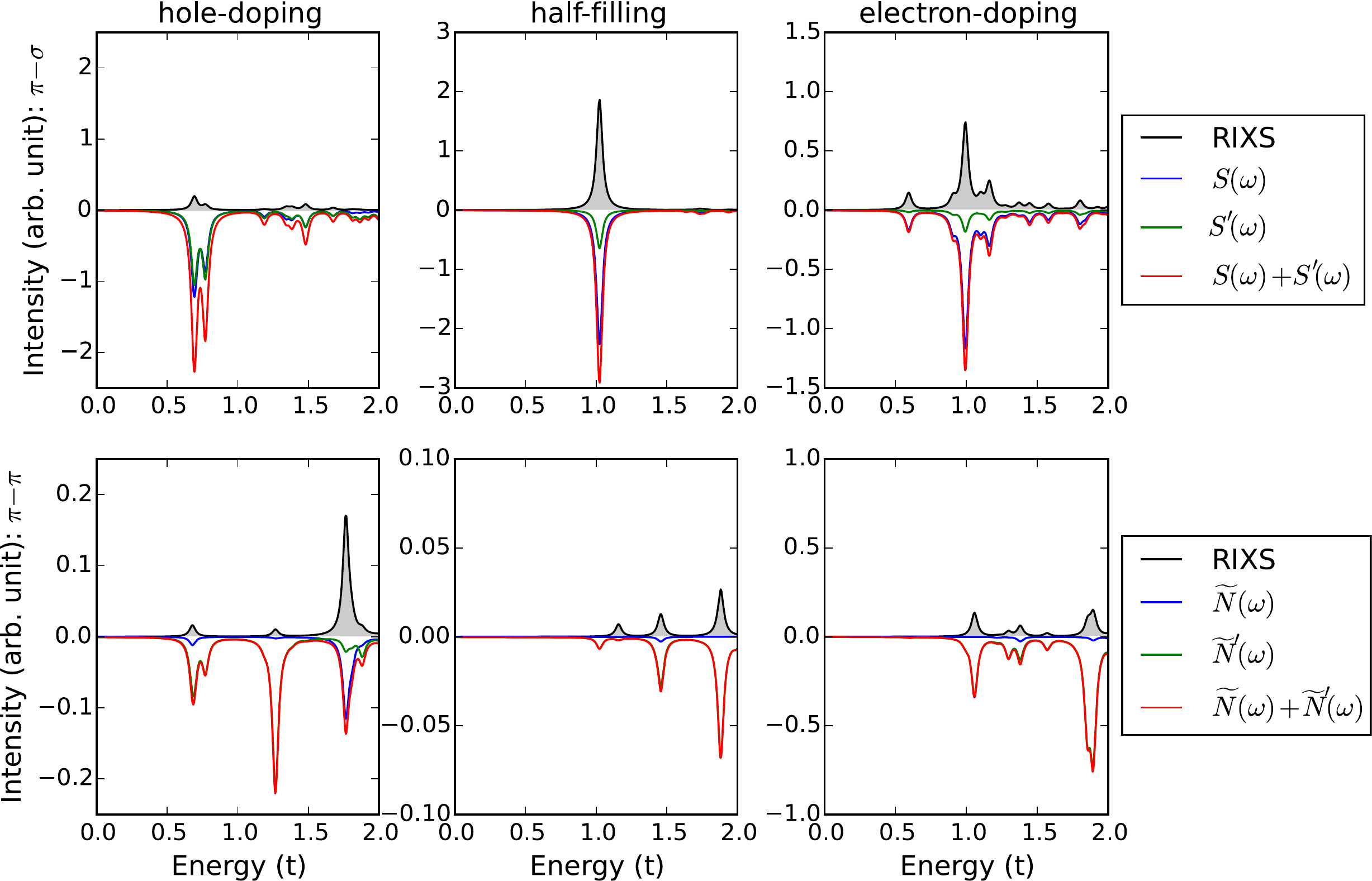}
 \caption{
Cross-sections at {\bf ${\bf q} = (\pi/2, \pi/2)$ point} with the elastic response removed.
Top (bottom) panels show spectra for the  $\pi-\sigma$ ($\pi-\pi$) polarization. 
Left, middle, and right panels show RIXS spectra calculated for $n=0.83$, 
$n=1$, 
and $n=1.17$ 
electron fillings, respectively. The approximate spectra are weighted with the RIXS form factors, i.e.
$S({\bf q}, \omega) \rightarrow  |{W}_{\bf \pi\!-\!\sigma} |^2 S({\bf q}, \omega) $, $\tilde{N} ({\bf q}, \omega) \rightarrow |{W}_{\bf \pi\!-\!\pi} |^2 \tilde{N} ({\bf q}, \omega)$
and similarly for $S' ({\bf q}, \omega)$ and  $\tilde{N'}({\bf q}, \omega)$.
}
\label{fig:spectrum_pi2}
\end{figure*}
%


First note the results in the cross-polarized channel (the $\pi-\sigma$ channel). On a qualitative level, the line shape of the full RIXS cross-section can be reproduced well by the spin dynamical structure factor $S({\bf q}, \omega)$
(the first term of the UCL approximation). 
This is true at half-filling, where all charge excitations have been gapped-out, 
while in either the electron- or hole-doped cases one can observe some relatively 
small 
discrepancies between the two spectra. 
When adding higher order terms from the effective expansion, to a large extent this observation remains unchanged 
since these terms encode similar excitations to $S({\bf q}, \omega)$ together with excitations of mixed charge and spin character, as can be seen readily from the form of the operator in Eq.~(8). 

However, on a quantitative level, 
this comparison breaks down, with 
discrepancies in the overall intensity and integrated spectral weight which 
can 
become very large 
(see Fig.~\ref{fig:spectralweight}). 
While one may have expected that higher order terms in the effective expansion should provide a more satisfactory qualitative and quantitative agreement, 
these do 
not help in reducing the 
differences; on the contrary, these additional terms 
actually enhance the quantitative mismatch. 
Note that the first two terms of the UCL expansion suggest larger spectral weight for the hole-doped case compared 
to that 
with electron-doping, 
in contrast to the behavior for the 
RIXS cross-section. 
This suggests that the differences 
cannot be attributed to a simple 
rescaling factor.  

The quantitative mismatch between RIXS at the Cu $L$-edge and the approximations highlights the role that the intermediate state wavefunction plays in the RIXS spectra, just as in the case for RIXS at the Cu $K$-edge.~\cite{Jia2012} RIXS is an intrinsic four-particle process, where the wavefunction overlap between the ground state and intermediate states, and between intermediate states and final states both play an important role. Neglecting the intermediate state wavefunction might still provide information on the fundamental excitation energies, but unfortunately it cannot provide reasonable spectral weights on a quantitative level. This mismatch also illustrates the failure of these approximations. That being said, both numerically and empirically, in the cross-polarized scattering geometry the RIXS cross-section qualitatively corresponds to the spin dynamical structure factor which encodes information about spin excitations at the two particle level, which underscores RIXS utility as a complementary probe to inelastic neutron scattering.

An altogether  
different situation arises in the parallel-polarized channel (the $\pi-\pi$ channel). While 
a comparison between the RIXS spectrum and 
the 
``projected'' charge excitations produces a modest qualitative agreement between the line shapes, 
both missing peaks as well as significant differences in the spectral weights undermine any quantitative value in this comparison.  The addition of the higher order terms 
seems to be needed for 
a 
better qualitative comparison of the line shapes 
although both terms support similar spectral excitations based on the form of the operators in Eqs.~(10) and (11) and a spectral weight analysis precludes any quantitative agreement.  In both cases, while the RIXS spectral lineshape may be approximated by the two expansion terms, neither provides a faithful representation for the proper two-particle charge response encoded in the simple dynamical structure factor, placing statements about the true charge excitation character of the RIXS cross-section on less solid footing. 

\subsection{Consequences for RIXS experiments} 
\label{sec:discussion}


The preceding section presented a comparison between cross-sections integrated in momentum, as well as energy for a total spectral weight analysis. In this section we show spectra at 
two particular momenta: ${\bf q} = (2\pi/3, 0)$ and ${\bf q} = (\pi/2, \pi/2)$ (see Fig.~\ref{fig:spectrum_2pi3} and Fig.~\ref{fig:spectrum_pi2}) to underscore those results, shown in a context amenable to experiment.  When doped, the Hubbard model, and by extension cuprates, will possess spin and charge excitations in a similar low energy regime which will appear, either directly or in a more complicated way reflecting the complexity of the cross-section, in the RIXS spectrum for the crossed- and parallel-polarization channels, respectively.  
Thus, to satisfactorily 
distinguish between the magnetic 
and charge channel, or two-spin excitations, one 
must perform 
measurements which can 
discriminate the outgoing polarizations. Unfortunately, to this point RIXS experiments have been unable to fully distinguish between the cross-polarized and parallel-polarized channels, making some statements with the help of a careful analysis of experimental RIXS scattering geometry. 
The newly constructed, 
state-of-the-art RIXS end-station at ESRF now provides an opportunity to perform such measurements (cf. Ref.~[\onlinecite{Minola2015}]); and other end-stations (currently operational or to be commissioned in the coming years) also would allow for the differentiation between the crossed- and parallel-polarized channels.

Even with 
outgoing 
polarization 
discrimination, in either the 
crossed- or parallel-polarized channels one needs to carefully 
invoke either $S(\mathbf{q},\omega)$ or $\tilde{N}(\mathbf{q},\omega)$ 
as 
approximations for the full RIXS cross-section. This 
especially may be true 
when analyzing 
RIXS spectral weights 
as a function of 
doping. 
None of these approximations address 
intermediate state 
effects, 
or equivalently differences in the cross-section with 
changes 
in the incoming photon energy.~\cite{Jia2012} Thus, to address the full resonant profile 
one always needs to calculate the full RIXS cross-section.
 

\section{Conclusions} 
\label{sec:conclusions}\
Overall, one must 
carefully 
apply 
approximations 
for calculating the RIXS cross-section. The nonlocal character of the intermediate state can become particularly important in 
correlated ground states with longer range entanglement. Therefore, we suggest the full RIXS simulations will be needed to verify the character of excitations. 

In the cross-polarized channel, we have shown that on a {\it qualitative} level 
Cu $L$-edge RIXS line shapes 
correspond to the spin dynamical structure factor $S({\bf q}, \omega)$, consistent with 
lowest order approximations 
as postulated by the fast collision approximation (or the effective operator approach)~\cite{Ament2009, Haverkort2010, Marra2012, Jia2014}, see Appendix C. 
As a consequence, we expect that the line shapes reported from cross-polarized RIXS experiments can be reproduced to some extent by theoretical modeling of the spin dynamical structure factors (or empirically through 
inelastic neutron scattering experiments when also considering differences in the effective matrix elements between the two techniques). However, the detailed analysis in this paper suggests that a {\it quantitative} comparison between RIXS and the two-particle spin and charge dynamical structure factors would be impractical. 
One should not expect a meaningful comparison between different spectral weights obtained from these different techniques, either experimentally or from simulation.

In the parallel-polarized channel, the situation is further 
complicated by the operator form taken by 
the approximations themselves. On a {\it qualitative} level the primary 
contributions 
seem to follow from higher order terms 
$\tilde{N'}({\bf q}, \omega)$, with a 
notable exception at 
half-filling, see Appendix C. 
However, any precise quantitative comparison to experiment would require calculating the full RIXS cross-section.  At the same time none of the terms in the approximations correspond to proper two-particle charge excitations, but rather inherently reflect the complexity of the RIXS process.  Thus, while 
line shapes 
in RIXS should closely resemble 
the line shapes of the (projected) charge excitations, 
the spectral weights may 
be quite different, with a more complicated analysis required to tease out the character of various spectral peaks.


\acknowledgments
We acknowledge support from the DOE-BES Division of Materials Sciences and Engineering (DMSE) under Contract No. DE-AC02-76SF00515 (Stanford/SIMES). 
K. W. acknowledges support from the Polish National Science Center (NCN) under Project No. 2012/04/A/ST3/00331. We are grateful for insightful discussions
with L. Braicovich, J. van den Brink, I.~Eremin, G. Ghiringhelli, D. J. Huang, B. J. Kim, A.~M.~Ole\'s, and G. A. Sawatzky.

\newpage

\beginsupplement

\section*{Appendix A: Derivation of the UCL approximation for RIXS at the Cu $L$-edge}
\label{app:ucl}


The UCL approximation should be valid when all the relevant eigenenergies of the intermediate state Hamiltonian $\mathcal{H}$ are much smaller than the inverse core-hole lifetime $\Gamma$. We use the following spectral decomposition
\begin{align}
\frac{1}{\omega_{i} - \mathcal{H} + i \Gamma} = \sum_{|N\rangle} |N \rangle \langle  N |
\frac{1}{\omega_{i} - {\mathcal{E}_N} + i \Gamma},
\end{align}
where $  \{ |N \rangle  \} $ are eigenstates of $\mathcal{H}$ with energy $ \{ \mathcal{E}_N \}$.
We are interested in RIXS at the resonant edge between the $2p^0 3d^{*1}$ initial configuration and the $2p^1 3d^{*0}$ intermediate state configuration. (All expressions are presented in hole language.) This means that we need to exclude all intermediate states $|N\rangle$
which contain one hole in the $d$ orbital on site ${\bf j}$, i.e. on the core-hole site. 
We add a projection operator $\tilde{P}_{\bf j}$ and rewrite the above expression as
\begin{align}
\sum_{|N\rangle} |N \rangle \langle  N |
\frac{1}{\omega_{i}-{\mathcal{E}_N} + i \Gamma} \rightarrow 
\sum_{|N\rangle} \tilde{P}_{\bf j} |N \rangle \langle  N | \tilde{P}_{\bf j}
\frac{1}{\omega_{i}-{\mathcal{E}_N} + i \Gamma}.
\end{align}

We define the following Hamiltonians 
$\bar{H} = H + \mathit{U}_c\sum_{{\bf i}\alpha\sigma\sigma^{\prime}} n_{{\bf i}\sigma}^d n_{{\bf i}\alpha\sigma^{\prime}}^p$
and
$\bar{H}_c = H_c - \mathit{U}_c\sum_{{\bf i}\alpha\sigma\sigma^{\prime}} n_{{\bf i}\sigma}^d n_{{\bf i}\alpha\sigma^{\prime}}^p$.
Note that $\bar{H}$ and $\bar{H}_c$ commute (and $\mathcal{H} = \bar{H} + \bar{H}_c$), since the intermediate states of
RIXS are such that they always contain either a hole in the $2p$ shell or in the $3d$ shell (guaranteed by the projection operators $\tilde{P}_{\bf j}$). Then we obtain:
\begin{align}
|N \rangle = |n \rangle |n_c \rangle  \quad \mathcal{E}_N = \varepsilon_n + \varepsilon_{n_c},
\end{align}
where $ |n \rangle$ are eigenstates of $\bar{H}$ with energy $\varepsilon_n$,
$ |n_c \rangle$ are eigenstates of $\bar{H}_c $ with energy $\varepsilon_{n_c}$.
Consequently we can write
\begin{align}
\frac{1}{\omega_{i} - \mathcal{H} + i \Gamma} = & \sum_{|n \rangle, |n_c \rangle} \tilde{P}_{\bf j} |n \rangle \langle  n | n_c \rangle \langle  n_c | \tilde{P}_{\bf j} \nonumber \\
& \times \frac{1}{\omega_{i}-\varepsilon_n - \varepsilon_{n_c} + i \Gamma}.
\end{align}

Note that with a single hole in the $p$ shell, there
are just two eigenstates of $\bar{H}_c$: $n_c \in {|L_2\rangle, |L_3 \rangle}$ with energies ${ \varepsilon_{L_2}, \varepsilon_{L_3}}$. 
They correspond to the two `$j$' eigenstates $j=1/2$ and $j=3/2$, respectively, split by the spin-orbit coupling $\propto \lambda$. (We note that $j$ represents the angular momentum and $\mathbf{j}$ represents site $\mathbf{j}$ on the cluster.) This implies
\begin{align}
\frac{1}{\omega_{i} - \mathcal{H} + i \Gamma} =&  \sum_{|n \rangle } \tilde{P}_{\bf j}  |n \rangle \langle  n | L_2 \rangle \langle  L_2 | \tilde{P}_{\bf j}
\frac{1}{\omega_{i}-\varepsilon_n - \varepsilon_{L_2} + i \Gamma} \nonumber \\
+ &  \sum_{|n \rangle}  \tilde{P}_{\bf j} |n \rangle \langle  n | L_3 \rangle \langle  L_3 | \tilde{P}_{\bf j}
\frac{1}{\omega_{i}-\varepsilon_n - \varepsilon_{L_3} + i \Gamma}.
\end{align}

{\it Resonance approximation}
We assume that the incoming x-ray photons are tuned to the $L_3$ resonance, i.e. $\omega_i \simeq \varepsilon_{L_3}$. Since $\lambda \gg \Gamma$ (and all eigenenergies $\varepsilon_n \ll \lambda$), we can neglect the contribution from the $L_2$ resonance and obtain
\begin{align}
\frac{1}{\omega_{i} - \mathcal{H} + i \Gamma} =  \sum_{|n \rangle}   \tilde{P}_{\bf j} |n \rangle \langle  n | L_3 \rangle \langle L_3 |  \tilde{P}_{\bf j} 
\frac{1}{-\varepsilon_n + i \Gamma}.
\end{align}

{\it UCL expansion}
We adopt the UCL expansion~\cite{Brink2006, Forte2008, Ament2010} for Cu $L$-edge RIXS to obtain 
\begin{align}
\frac{1}{\omega_{i} - \mathcal{H} + i \Gamma} = \tilde{P}_{\bf j} |L_3 \rangle \langle L_3 | 
\sum_{l=0}^{+\infty} \frac{{\bar{H}}^l}{(i \Gamma)^{l+1}} \tilde{P}_{\bf j}.
\end{align}
This means that the RIXS operator can be rewritten as
\begin{align}
{O}_{{\bf j}, {\bf e}} = \frac{1}{i \Gamma} D_{{\bf j}, {\bf e}^f}^\dag \tilde{P}_{\bf j}
 |L_3 \rangle \langle L_3| \sum_{l=0}^{+\infty} \frac{{\bar{H}}^l}{(i \Gamma)^{l}} \tilde{P}_{\bf j}
D_{{\bf j}, {\bf e}^i}.
\end{align}

{\it Second order UCL approximation}
Keeping only terms with $l=0$ and $l=1$ gives
\begin{align}
{O}_{{\bf j}, {\bf e}} = & {O}^{(1)}_{{\bf j}, {\bf e}} + {O}^{(2)}_{{\bf j}, {\bf e}}  \\ \nonumber
{O}^{(1)}_{{\bf j}, {\bf e}} =
& \frac{1}{i \Gamma} D_{{\bf j}, {\bf e}^f}^\dag \tilde{P}_{\bf j}
 |L_3 \rangle \langle L_3|  \tilde{P}_{\bf j} D_{{\bf j}, {\bf e}^i} \nonumber \\
{O}^{(2)}_{{\bf j}, {\bf e}} =
& \frac{1}{(i \Gamma)^2}  D_{{\bf j}, {\bf e}^f}^\dag \tilde{P}_{\bf j}
|L_3 \rangle \langle L_3| {\bar{H}} \tilde{P}_{\bf j} D_{{\bf j}, {\bf e}^i}.
\end{align}
The validity of this approximation has been discussed in detail in the main text of the paper.

{\it Change of projection operators}
It is convenient to use another 
operator -- $P_{\bf j}$ -- which projects to the sector with no double
occupancy in the $d$ level on site ${\bf j}$, i.e. where the core-hole is created. This gives
\begin{align}
\tilde{P}_{\bf j} D_{{\bf j}, {\bf e}^i}  = D_{{\bf j}, {\bf e}^i}  {P}_{\bf j}.
\end{align}
A similar expression holds also for the $D^{\dag}_{{\bf j}, {\bf e}^f}$ dipole operator.

{\it The $l=1$ term} Next we evaluate 
\begin{align}
& P_{\bf j} D_{{\bf j}, {\bf e}^f}^\dag |L_3 \rangle \langle L_3|  {\bar{H}} D_{{\bf j}, {\bf e}^i} P_{\bf j} |i \rangle. 
\end{align}
Since
\begin{align}
P_{\bf j} D_{{\bf j}, {\bf e}^f}^\dag |L_3 \rangle \langle L_3| \bar{H}_{\bf mj} D_{{\bf j}, {\bf e}^i} P_{\bf j} |i \rangle =0
\end{align}
where 
\begin{align}
\bar{H}_{\bf mj} =& -t\! \! \sum_{m(j) , \sigma} ({d}^\dag_{{\bf m} \sigma} {d}_{{\bf j} \sigma}\! +\! h.c.)
-t' \! \! \sum_{{\bf m}'({\bf j}) , \sigma} ({d}^\dag_{{\bf m}' \sigma} {d}_{{\bf j} \sigma} \!+\! h.c.) \nonumber \\ 
&+\mathit{U}_c\sum_{\alpha\sigma\sigma^{\prime}} n_{{\bf j} \sigma}^d n_{{\bf j} \alpha\sigma^{\prime}}^p+ \!U\!n_{{\bf j} \uparrow}^d n_{{\bf j} \downarrow}^d 
\end{align}
due to the fact that all terms in the Hamiltonian which contain site {\bf j} will vanish when ``sandwiched''' between
the dipole operators $D$ and evaluated on the initial state $| i \rangle$. Here ${\bf m}({\bf j})$ and ${\bf m}'({\bf j})$ are nearest and next nearest neighbors of site ${\bf j}$.

Thus, the following expression holds:
\begin{align}\label{eq:expr}
&P_{\bf j} D_{{\bf j}, {\bf e}^f}^\dag |L_3 \rangle \langle L_3|  {\bar{H}} D_{{\bf j}, {\bf e}^i} P_{\bf j} |i \rangle = 
P_{\bf j} D_{{\bf j}, {\bf e}^f}^\dag |L_3 \rangle \langle L_3| \nonumber \\
&\times ( {\bar{H}} - \bar{H}_{\bf mj}) D_{{\bf j}, {\bf e}^i} P_{\bf j} |i \rangle.
\end{align}
Commuting the Hamiltonian ${\bar{H}} - \bar{H}_{\bf mj}$ (which does not contain operators on site {\bf j}) with the operator $D_{{\bf j}, {\bf e}^i} $ and $P_{\bf j}$ to gives
\begin{align}
&P_{\bf j} D_{{\bf j}, {\bf e}^f}^\dag |L_3 \rangle \langle L_3|  {\bar{H}} D_{{\bf j}, {\bf e}^i} P_{\bf j} |i \rangle = 
P_{\bf j} D_{{\bf j}, {\bf e}^f}^\dag |L_3 \rangle \langle L_3| \nonumber \\
& \times D_{{\bf j}, {\bf e}^i} P_{\bf j} ( {\bar{H}} - \bar{H}_{\bf mj})  |i \rangle.
\end{align}
Since $\bar{H} | i \rangle =H | i \rangle= E_i | i \rangle =0$ to set the energy zero, we are left with the following expression $ P_{\bf j} D_{{\bf j}, {\bf e}^f}^\dag |L_3 \rangle \langle L_3| D_{{\bf j}, {\bf e}^i} P_{\bf j} \bar{H}_{\bf mj}  |i \rangle$. However, due to $P_{\bf j}$, the $U$ and the $U_c$ terms will never contribute and $P_{\bf j} D_{{\bf j}, {\bf e}^f}^\dag |L_3 \rangle \langle L_3| D_{{\bf j}, {\bf e}^i} P_{\bf j} {d}^\dag_{{\bf l} \sigma} {d}_{{\bf j} \sigma}) |i \rangle =0$ for any ${\bf l}={\bf m}, {\bf m}'$.
Thus, we obtain
\begin{align}
&P_{\bf j} D_{{\bf j}, {\bf e}^f}^\dag |L_3 \rangle \langle L_3|  {\bar{H}} D_{{\bf j}, {\bf e}^i} P_{\bf j} |i \rangle = 
-P_{\bf j}  D_{{\bf j}, {\bf e}^f}^\dag |L_3 \rangle \langle L_3| D_{{\bf j}, {\bf e}^i}\nonumber \\ 
& \times  [-t \sum_{{\bf m}({\bf j}), \sigma} ({d}^\dag_{{\bf j} \sigma} {d}_{{\bf m} \sigma})
-t' \sum_{{\bf m}'({\bf j}), \sigma} ({d}^\dag_{{\bf j} \sigma} {d}_{{\bf m} \sigma})] P_{\bf j} |i \rangle.
\end{align}
Note the asymmetry in the above expression, i.e. the lack of the hermitian conjugate terms $\propto {d}^\dag_{j \sigma} {d}_{m \sigma}$
 -- this asymmetry expresses the fact that in RIXS we are only sensitive to sites on which the $3d$ holes reside.

Introducing so-called local matrix elements of RIXS, we obtain
\begin{align}
\frac{1}{i \Gamma} P_{\bf j} D_{{\bf j}, {\bf e}^f}^\dag 
|L_3 \rangle \langle L_3|  D_{{\bf j}, {\bf e}^i} P_{\bf j}  \equiv W_{\bf e} P_{\bf j} {n}_j P_{\bf j} + \tilde{W}_{\bf e} P_{\bf j} S^z_{\bf j} P_{\bf j},
\end{align}
where the local RIXS form factors follow from e.g. Ref. [\onlinecite{Marra2012}] (cf. Eq. (2) and Fig. 1 in Ref. [\onlinecite{Marra2012}]):
\begin{align}
{W}_{\bf \pi\!-\!\sigma} & =- \imath 2 ({\bf e}^i_y {\bf e}^f_x - {\bf e}^i_x {\bf e}^f_y) / (3 \Gamma), \nonumber \\
{W}_{\bf \pi\!-\!\pi}  &=-  2 ({\bf e}^i_x {\bf e}^f_x + {\bf e}^i_y {\bf e}^f_y) / (3 \Gamma).
\end{align}
Let $P_{\bf j} {n}_{\bf j} P_{\bf j} = \tilde{n}_{\bf j}$ [where $\tilde{n}_{\bf j} =  \sum_{\sigma} \tilde{n}_{\bf j \sigma} = \sum_{\sigma} \tilde{d}^{\dag}_{{\bf j} \sigma} \tilde{d}_{{\bf j} \sigma}$ and $ \tilde{d}^{\dag}_{{\bf j} \sigma} ={d}^{\dag}_{{\bf j} \sigma}  (1- {n}_{ {\bf j}, -\sigma})$] and $P_{\bf j} S^z_{\bf j} P_{\bf j} = S^z_{\bf j}$. Combining the above equations, we finally arrive at the expression for the RIXS operators in the UCL approximation
\begin{align}
{O}^{(1)}_{{\bf j}, {\bf e}} =& W_{\bf e} \tilde{n}_{\bf j} + \tilde{W}_{\bf e} S^z_{\bf j}, \nonumber \\
{O}^{(2)}_{{\bf  j}, {\bf e}} = & \frac{t}{i \Gamma} W_{\bf e} \sum_{{\bf m} ({\bf j})} \tilde{d}^\dag_{{\bf j} \sigma} {d}_{{\bf m} \sigma}
 + \frac{t}{i \Gamma} \tilde{W}_{\bf e} S^z_{\bf j} \sum_{{\bf m} ({\bf j})} {d}^\dag_{{\bf j} \sigma} {d}_{{\bf m} \sigma} \nonumber \\
&+ \frac{t'}{i \Gamma} W_{\bf e} \sum_{{\bf m'} ({\bf j})} \tilde{d}^\dag_{{\bf j} \sigma} {d}_{{\bf m} \sigma}
 + \frac{t'}{i \Gamma} \tilde{W}_{\bf e} S^z_{\bf j} \sum_{{\bf m}' ({\bf j})} {d}^\dag_{{\bf j} \sigma} {d}_{{\bf m} \sigma}.
\end{align}

Substituting the above expressions into Eq.~(\ref{eq:crosssection}) and performing Fourier transformations, we obtain Eqs.~(\ref{eq:uclpisigma}) and (\ref{eq:uclpipi}). As the first (second) order UCL terms have real (imaginary) contributions, the interference terms vanish and the full RIXS cross-section consists of separate first and second order UCL terms. Note that if only these first two terms are considered in the UCL approximation, then the RIXS spectrum does not depend on the size of the core-hole potential $U_c$ (the latter will appear only in higher order corrections in the UCL approximation).

\section*{Appendix B: UCL approximation for the $t-J$ model}
\label{sec:approximate2}

For completeness, we have evaluated 
the UCL expansion of the RIXS cross-section for the $t$--$J$ model -- the strong coupling expansion of the Hubbard model, valid for the low energy physics well below the energy scale $U=8t$. For the qualitative discussions here, we can safely neglect $t'$ and the three-site terms in this expansion. Following similar steps as described in the previous section for the Hubbard model, we obtain for the $\pi-\sigma$ channel:
\begin{align}
&I^{\rm UCL_{t-J}}_{\pi-\sigma}({\bf q}, \omega)=   |{W}_{\pi-\sigma} |^2  \Big\{
\sum_f \Big| \Big\langle f \Big\vert  S^z_{\bf q} \Big\vert i \Big\rangle \Big|^2 \delta (\omega + E_i - E_f) \nonumber \\
&+ \frac{z^2 J^2}{N^2 \Gamma^2} \sum_f \Big| \Big\langle f \Big\vert  
\sum_{{\bf k }, {\bf k'}} \gamma_{{\bf k'}+ {\bf k}-{\bf q}}  S^z_{\bf k'} {\bf S}_{\bf k} {\bf S}_{-{\bf k'}-{\bf k}+{\bf q}} \Big\vert i \Big\rangle \Big|^2  \nonumber \\
&\times \delta (\omega + E_i - E_f) 
\nonumber \\
& + \frac{z^2 t^2}{N^2 \Gamma^2} \sum_f \Big| \Big\langle f \Big\vert   \sum_{{\bf k}, {\bf k'}} S^z_{\bf k'} \gamma_{-{\bf k'}+{\bf k}+{\bf q}} 
\tilde{d}_{{\bf k}, \sigma}^\dag \tilde{d}_{-{\bf k'}+{\bf k}+{\bf q}, \sigma} \Big\vert i \Big\rangle \Big|^2 \nonumber \\
&\times \delta (\omega + E_i - E_f)  \Big\},
\end{align}
and for the $\pi-\pi$ channel:
\begin{align}
&I^{\rm UCL_{t-J}}_{\pi-\pi}({\bf q}, \omega)=   |{W}_{\pi-\pi} |^2  \Big\{
\sum_f \Big| \Big\langle f \Big\vert  \tilde{n}_{\bf q} \Big\vert i \Big\rangle \Big|^2 \delta (\omega + E_i - E_f) \nonumber \\
&+ \frac{z^2 J^2}{ N \Gamma^2}\sum_f \Big| \Big\langle f \Big\vert  \sum_{\bf k} \gamma_{{\bf q}-{\bf k}} {\bf S}_{\bf k} {\bf S}_{-{\bf k}+{\bf q}}  \Big\vert i \Big\rangle \Big|^2 \delta (\omega + E_i - E_f) \nonumber \\
&+ \frac{z^2 t^2}{N \Gamma^2} \sum_f \Big| \Big\langle f \Big\vert  \sum_{\bf k} \gamma_{{\bf q}+{\bf k}} \tilde{d}_{{\bf k}, \sigma}^\dag \tilde{d}_{{\bf q}+{\bf k}, \sigma} \Big\vert i \Big\rangle \Big|^2  \nonumber \\
&\times \delta (\omega + E_i - E_f) \Big\}.
\end{align}
Here $J=4t^2 / U$ and the spin operators are defined in a standard way as 
 \begin{align}
{\bf S}_{\bf k} {\bf S}_{-{\bf q}-{\bf k}}  = & \frac{1}{2 {N}}  \sum_{{\bf q}_1, {\bf q}_2, \sigma} \tilde{d}_{{\bf q}_1, \sigma}^\dag \tilde{d}_{{\bf q}_1+{\bf q}, \bar{\sigma}}  
\tilde{d}_{{\bf q}_2, \bar{\sigma}}^\dag \tilde{d}_{{\bf q}_2-{\bf q}-{\bf k}, {\sigma}} \nonumber \\
 &+ \frac{1}{4 L} \sum_{{\bf q}_1, {\bf q}_2} (\tilde{d}_{{\bf q}_1, \uparrow}^\dag \tilde{d}_{{\bf q}_1+{\bf k}, \uparrow}  -
\tilde{d}_{{\bf q}_1, \downarrow}^\dag \tilde{d}_{{\bf q}_1+{\bf k}, \downarrow}) \nonumber \\
&(\tilde{d}_{{\bf q}_2, \uparrow}^\dag \tilde{d}_{{\bf q}_2-{\bf q} -{\bf k}, \uparrow}  -
\tilde{d}_{{\bf q}_2, \downarrow}^\dag \tilde{d}_{{\bf q}_2-{\bf q} - {\bf k}, \downarrow}). 
\end{align}
Note that all the operators $\tilde{d}_{{\bf i}, \sigma}$ and  $\tilde{d}^{\dag}_{{\bf i}, \sigma}$ are defined in the constrained Hilbert space without double occupancies.

\begin{figure}[t!]
\includegraphics[width=1.0\columnwidth]{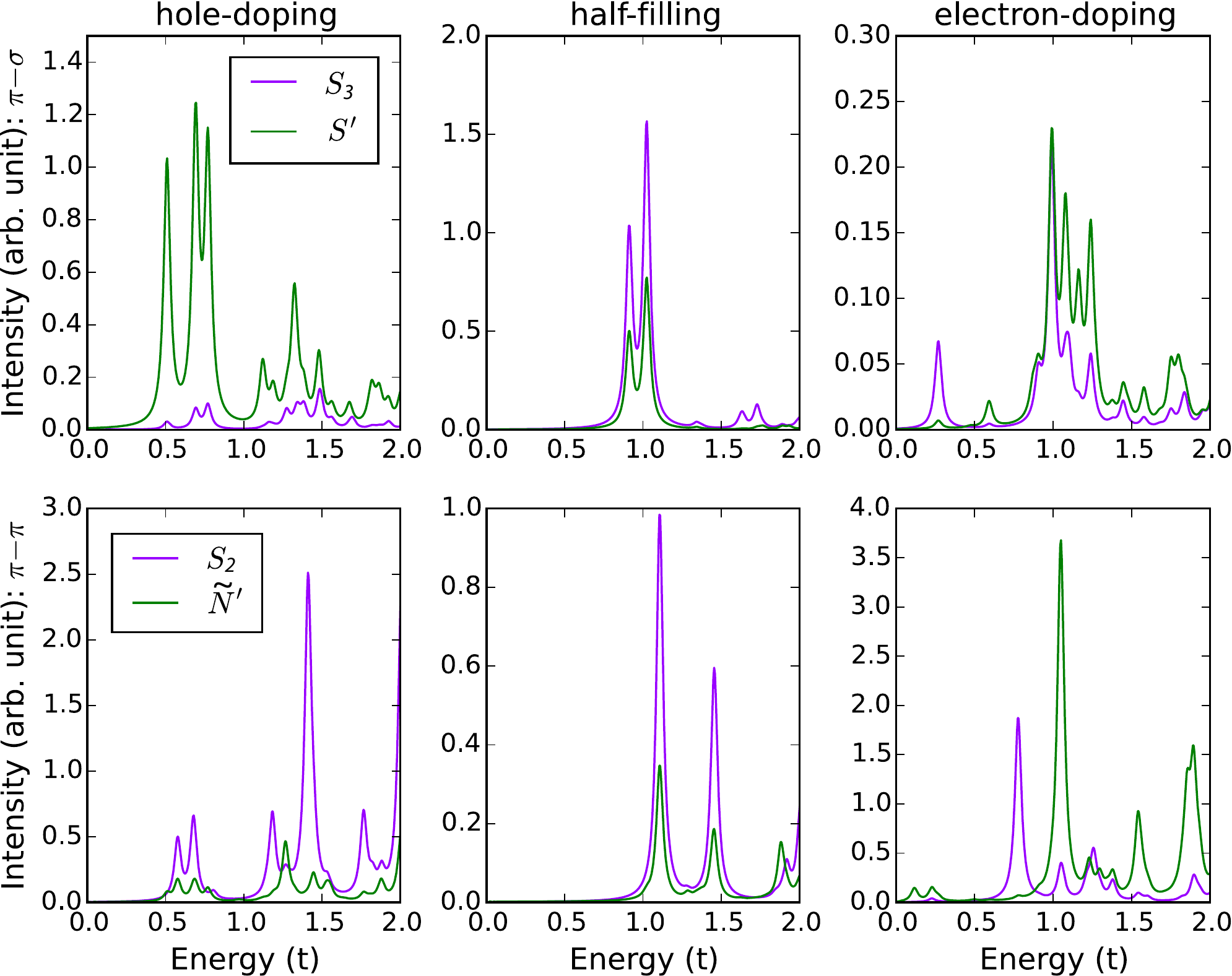}
 \caption{Comparison between ${S'}$ and 3-spin Green's functions $S_3$ and between 
$\tilde{N'}$ and 2-spin Green's functions $S_2$ calculated using 
exact diagonalization and following the Eqs. (\ref{eq:tildeS}) and (\ref{eq:tildeN}) from the main text of the paper and
Eqs. (\ref{eq:s2}) and (\ref{eq:s3}). Top (bottom) panels show spectra for the  $\pi-\sigma$ ($\pi-\pi$) polarization setups, see 
main text of the paper for further details.
Left, middle, and right panels show RIXS spectra calculated for $n=0.83$, 
$n=1$, 
and $n=1.17$ 
electron fillings, respectively. The spectra are summed over all
momenta along the nodal direction and along the $k_y=0$ direction that are
accessible in RIXS and on a 12-site cluster used in the exact diagonalization calculations and weighted with the RIXS form factors, i.e.:
$S' (\omega)= \sum_{\bf q}  |{W}_{\bf \pi\!-\!\sigma} |^2 {S'}({\bf q}, \omega) $, 
$\tilde{N'}(\omega)= \sum_{\bf q}  |{W}_{\bf \pi\!-\!\pi} |^2 \tilde{N'}({\bf q}, \omega)$,
$S_2 (\omega) = \sum_{\bf q} |{W}_{\bf \pi\!-\!\pi} |^2  S_{2}({\bf q}, \omega)$, and
$S_3 (\omega) = \sum_{\bf q} |{W}_{\bf \pi\!-\!\sigma} |^2  S_{3}({\bf q}, \omega)$.
The intensity scale is different on each panel (it is chosen in such a way that each of the 6 spectra can be well visible) and the elastic response has been removed.
}
\label{fig:23mag}
\end{figure}

These equations show that in the second order of the UCL expansion we obtain two terms. 
The first ones contain only the spin operators and correspond to the `2-spin'
\begin{align}
\label{eq:s2}
S_{2}({\bf q}, \omega)=& \frac{z^2 J^2}{ N \Gamma^2}\sum_f \Big| \Big\langle f \Big\vert  \sum_{\bf k} \gamma_{{\bf q}-{\bf k}} {\bf S}_{\bf k} {\bf S}_{{\bf k}-{\bf q}}  \Big\vert i \Big\rangle \Big|^2 \nonumber \\ 
& \times \delta (\omega + E_i - E_f),
\end{align}
and `3-spin' Green's functions
\begin{align}
\label{eq:s3}
S_{3}({\bf q}, \omega)=& \frac{z^2 J^2}{L^2 \Gamma^2} \sum_f \Big| \Big\langle f \Big\vert  
\sum_{{\bf k }, {\bf k'}} \gamma_{{\bf k'}+ {\bf k}-{\bf q}}  S^z_{\bf k'} \nonumber \\ 
& \times {\bf S}_{\bf k} {\bf S}_{-{\bf k'}-{\bf k}+{\bf q}} \Big\vert i \Big\rangle \Big|^2  
\delta (\omega + E_i - E_f).
\end{align}
The second terms always involve charge excitations below the gap. Since these terms do not contribute at half-filling (due to the constrained Hilbert space without double occupancies), it is expected that at relatively low doping levels, the first terms should be dominant (even though their amplitude scales with $J/\Gamma$ and not with $t/\Gamma$). Hence, we compare the spectra of these first two terms, $S_{2}({\bf q}, \omega)$ and $S_{3}({\bf q}, \omega)$, in the second order of the UCL expansion with the spectra of $S' ({\bf q}, \omega)$ and $\tilde{N'} ({\bf q}, \omega)$ (cf. Fig.~\ref{fig:23mag}) from the Hubbard model. We see that at half-filling $S' ({\bf q}, \omega)$ can be approximated relatively well by the `3-spin' excitations probed by $S_{3}({\bf q}, \omega)$ and that $\tilde{N'} ({\bf q}, \omega)$ can be approximated relatively well by 
the ``2-spin'' or bi-magnon excitations probed by $S_{2}({\bf q}, \omega)$. Most of the discrepancies between these two spectra can be found in the high energy regime and are therefore attributed to the failure of the $t$--$J$ model expansion at higher energies.  The electron- and hole-doped cases show much less pronounced agreement, where the projected spin and charge excitations $S' ({\bf q}, \omega)$ and $\tilde{N'} ({\bf q}, \omega)$ also probe the low energy charge excitations below the gap which can give a relatively large contribution in the spectrum.

The prediction that RIXS can probe the ``2-spin'' and the ``3-spin'' excitations at half-filling already has been put forward in 
Refs.~\cite{Ament2010, Bisogni2012a, Bisogni2012b, Kourtis2012, Igarashi2012a, Igarashi2012b} in the case of the Heisenberg model, consistent with our UCL expansions for the Hubbard and $t$--$J$ models.  Note that usually the Greens functions containing the `2-spin' and the `3-spin' operators are referred to as probing the `two-magnon' and the `three-magnon' spectrum, though this terminology may be used loosely in this context.  Finally, the fact that the charge dynamical structure factor for the half-filled Hubbard model probes the `two-magnon' spectrum also has been discussed in the context of nonresonant Raman scattering (cf. Ref.~[\onlinecite{Morr1996, Devereaux2007, Lin2012}]).

\section*{Appendix C: Visualization of the final state configurations}

\begin{figure*}[t!]
\includegraphics[width=1.8\columnwidth]{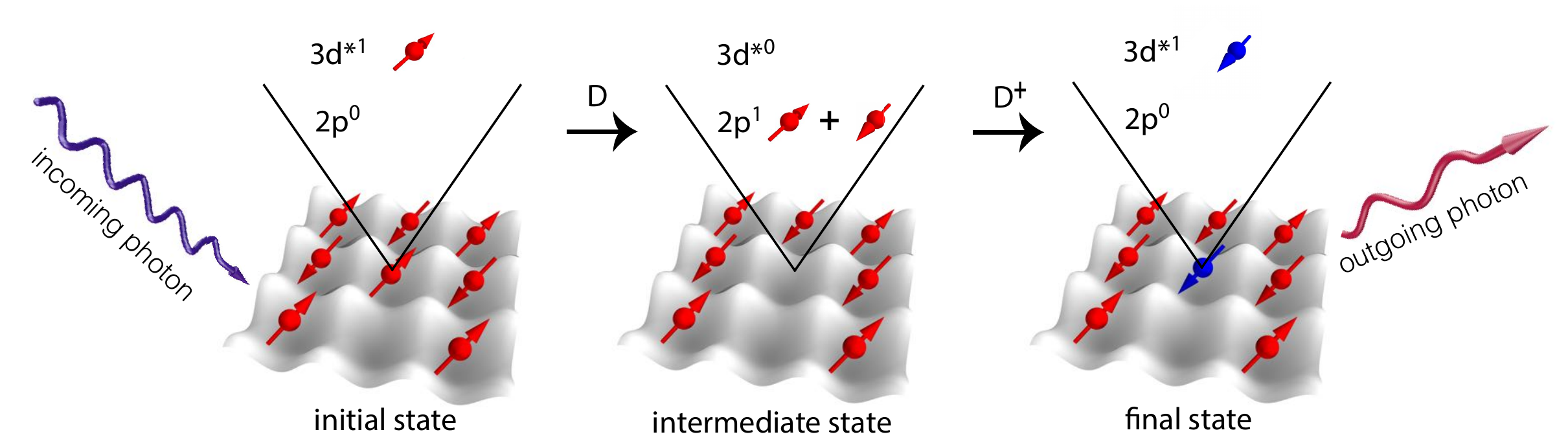}
 \caption{A cartoon picture of the dominant Cu $L$-edge RIXS process in the cross-polarized `channel'. 
}
\label{fig:cartoonpisigma}
\end{figure*}

Fig.~\ref{fig:cartoonpisigma} shows the dominant Cu $L$-edge RIXS process in the cross-polarized channel in hole language. The hole in the $3d^*$ orbital on a particular site ${\bf j}$ in the initial state of RIXS is transferred via the dipole operator $D$ into the $2p$ orbital on the same site ${\bf j}$ with a 
non-uniquely defined spin in the intermediate state of RIXS due to spin-orbit coupling in the core. This is transferred back via the dipole operator $D^{\dag}$ to the $d$ orbital on the same site ${\bf j}$ with a spin flip in the final state compared to the initial state of the RIXS process.  While 
we demonstrate this process on a single site ${\bf j}$ in real space, in reality a coherent superposition of such excitations are created with phase factors $e^{\imath {\bf q} {\bf j}}$; 
this leads to a well-defined, single spin flip with momentum ${\bf q}$ in the final state of RIXS, i.e. RIXS is sensitive to the spin dynamical structure factor ${S}({\bf q}, \omega)$.

\begin{figure*}[ht!]
\includegraphics[width=1.8\columnwidth]{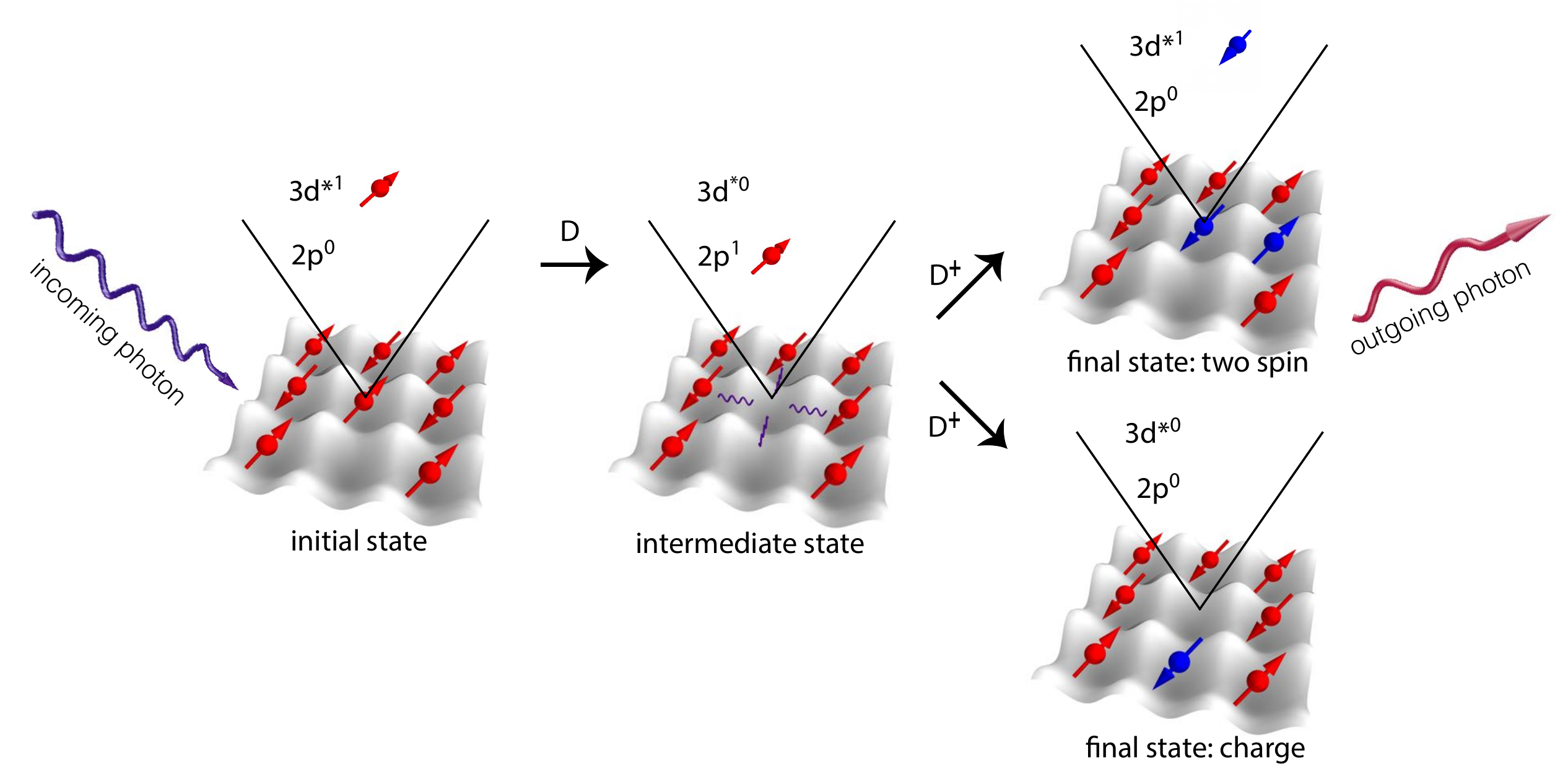}
 \caption{A cartoon picture of the dominant Cu $L$ edge RIXS process in the parallel-polarized `channel'. 
}
\label{fig:cartoonpipi}
\end{figure*}

Fig.~\ref{fig:cartoonpipi} shows the dominant Cu $L$-edge RIXS process in the parallel-polarized channel in hole language. The hole in the $3d^*$ orbital on a particular site ${\bf j}$ in the initial state of RIXS is transferred via the dipole operator $D$ into the $2p$ orbital on the same site ${\bf j}$ with a well defined spin in the intermediate state. 
In the intermediate state 
a ``shakeup'' 
happens, which creates a ``2-spin'' and/{\it or} charge excitation in the final state. 
While this process is shown 
on a single site ${\bf j}$ (and its neighbors) in real space, 
in reality a coherent superposition of such excitations is created with a phase factor $e^{\imath {\bf q} {\bf j}}$; 
leading to a ``2-spin'' or charge excitation created with momentum ${\bf q}$ in the final state, i.e. RIXS has some sensitivity to $\tilde{N'}({\bf q}, \omega)$, which unfortunately have no simple analog in the standard two-particle charge response function.

%

\bibliographystyle{prsty-etal}
\bibliography{Rixs}
\end{document}